\documentclass{article}
\usepackage[totalwidth=450pt, totalheight=600pt]{geometry}

\usepackage{ifthen}
\usepackage[latin1]{inputenc}
\usepackage{amsmath,amssymb,amsthm}
\def\NN{{\mathbb N}}

\def\QQ{{\mathbb Q}}
\def\RR{{\mathbb R}}

\def\ZZ{{\mathbb Z}}
\newcommand{\fd}{\ensuremath{\rightarrow}}

\newcommand{\findem}{\nolinebreak\vspace{\baselineskip} \hfill\rule{2mm}{2mm}}
\renewcommand{\phi}{\ensuremath{\varphi}}
\newcommand{\inc}{\ensuremath{\subset}}

\renewcommand{\phi}{\ensuremath{\varphi}}
\newcommand{\x}{\ensuremath{\times}}

\newtheorem{nt}{Notation}

\newtheorem{prop}[nt]{Proposition}

\newtheorem{exercice}{Exercice} 
\newcounter{numeroquestion} 
\newcounter{numerosousquestion}

\newcommand{\sousquestion}{\ifthenelse{\value{numerosousquestion}=1}{}{\\}\textbf{\roman{numerosousquestion})} \addtocounter{numerosousquestion}{1}}
\newtheorem{correction}{Correction}

\newenvironment{listecompacte}
{\begin{list}
    {\ensuremath{\bullet}}
    {\setlength{\topsep}{2pt}
      \setlength{\itemsep}{1pt} \setlength{\parsep}{0pt}}
}
{\end{list}
}

\newtheorem{coro}[nt]{Corollary} 
\newtheorem{defi}[nt]{Definition}

\newtheorem{lm}[nt]{Lemma} 
 
\newtheorem{rem}[nt]{Remark} 
\newtheorem{thm}[nt]{Theorem}

\newenvironment{dem}{\noindent\textit{Proof.} }{\findem}

\begin{document}
\sloppy
\title{Knapsack cryptosystems built on NP-hard instances}
\date{}
\author{Laurent Evain (laurent.evain@univ-angers.fr)}
\maketitle

%%%%%%%%%%%%%%%%%%%%%%%%%%%%%%%%%%%%%%%%%%%%%%%%%%%%%%%%%%%%%%%%
%%%%                    Debut du contenu                  %%%%%
%%%%%%%%%%%%%%%%%%%%%%%%%%%%%%%%%%%%%%%%%%%%%%%%%%%%%%%%%%%%%%%%

\section*{Abstract: }
We construct three public key knapsack cryptosystems. 
Standard knapsack cryptosystems hide easy instances
of the knapsack problem and have been broken. The systems considered in
the article face this problem: They 
hide a random (possibly hard) instance of the knapsack
problem. We provide both complexity results (size of 
the key, time needed to encypher/decypher...)
and experimental results.
Security results are given for the second cryptosystem
( the fastest one and the one with the shortest key). 
Probabilistic polynomial reductions show that 
finding the private key is as difficult 
as factorizing a product of two primes. We also consider heuristic
attacks. First, the density of the cryptosystem can be chosen
arbitrarily close to one, discarding low density attacks. 
Finally, we consider explicit heuristic attacks based on the LLL algorithm and we
prove that with respect to these attacks, the public key 
is as secure as a random key.

\section*{Introduction}
\label{sec:introduction}

\subsection*{The principle}
\label{sec:principle}

It is natural to build cryptosystems relying on NP-complete
problems since NP-complete problems are presumably difficult to
solve. There are several versions of knapsack problems, all of them being
NP-complete. Several cryptosystems relying 
on knapsack problems have been introduced in the eighties \cite{odlyzko89:cryptoSurvey}

 We are interested in the bounded version of the knapsack problem. Let 
$s,M,v,v_1,\dots , v_s \in \NN$. The problem is to determine
whether there are integers 
$\epsilon_i$, $0\leq \epsilon_i <M$ such that $\sum_{i=1}^{i=s} \epsilon_i
v_i=v$. In case $M=2$, the problem is to 
fill a knapsack of volume $v$
with objects of  volume $v_i$. 

Knapsack cryptosystems are built on knapsack problems. 
Alice constructs integers $v_i$ (using some private key $q$)
such that the cyphering map $C$ is injective: $C:\{0,\dots,M-1\}^s \fd \NN$, $(\epsilon_i)\mapsto
\sum \epsilon_i v_i$. The sequence $v_i$ is the public
key. When Bob has a plaintext message $m\in 
\{0,\dots,M-1\}^s$ for Alice, he sends the ciphertext $C(m)$. 
Alice decodes using her private key.

\subsection*{Strength and weakness of knapsack cryptosystems}
\label{sec:advant-inconv}

The main advantage of knapsack cryptosystems is the speed.
These systems attain very high encryption and decryption rates.  
The knapsack cryptosystem proposed by Merkle-Hellman 
\cite{merkleHellman82} 
seemed to be 100 times faster than RSA for the same level of
security at the time it was introduced \cite{odlyzko89:cryptoSurvey}. 

The main weakness of knapsack cryptosystems is security.
% The Merkel-Hellman system has been cryptanalysed. 
% Shamir announced a theoretical polynomial attack in 1982 \cite{} and
% implementations have followed \cite{}. 
% Several modifications of the scheme
% have been proposed, but the variations ended up with 
% Brickell's attack on the
% iterated Merkel-Hellmann cryptosystem. Brickell's attack was an evidence that the 
%Merkel-Hellman cryptosystem could not be made secure with
%slight modifications. 
%Other 
All standard knapsack cryptosystems have been broken:
the Merkle-Hellman cryptosystem by Shamir and Adleman
\cite{shamir82:cassageMerkleHelman},
, the iterated
Merkle-Hellmann by Brickell 
\cite{brickell85:cassageKnapsackItere}
, the Chor-Rivest cryptosystem by
Vaudenay in 1997 
\cite{vaudenay98:cassageChorRivest}
...

Two main reasons explain the fragility of knapsack
cryptosystems. 

First, most of these cryptosystems start with 
an easy instance.
The knapskack problem is
NP-complete and no fast algorithm to solve it is known in general. However, the knapsack 
problem is 
easy to solve for some instances $(v_i)_{i\leq s}$:
%if $v_i=2^i$, then the $\epsilon_i$'s are 
%the coefficients of the diadic representation of $v$.
%More generally, 
if $(v_i)$ is a superincreasing sequence in the sense that
$v_i> \sum_{j<i} v_j$, there is a very fast algorithm to solve the 
knapsack problem, depending linearly on the size of the data.
For knapsack cryptosystems, the public key is usually a hard instance $(v_i)$ obtained 
as a function  $v_i=f(q,w_i)$ of an easy
instance $(w_i)$ using a private key $q$. 
% For instance, 
% Merkle-Hellmann start with a superincreasing sequence $(w_i)$ and the 
% integers $w_i$ are considered as elements 
% in $\ZZ/p\ZZ$ for $p>>0$. The private key of Alice is an
% element $q$ invertible in the ring $\ZZ/p\ZZ$. The public key 
% is the sequence $(v_i)=(qw_i\ \mod\ p)$. When Alice receives a
% ciphertext encrypted with the hard instance $(qw_i) \mod p$,  she
% multiplies it by $q^{-1}\ \mod p$ to
% obtain the cyphertext associated with
% the easy instance $(w_i)$. 
When Alice receives the message $C_{v_i}(m)$ encrypted with the hard instance $v_i$,  she
can compute with her private key the message $C_{w_i}(m)$ encrypted with 
the easy instance $w_i$. Then she decodes easily. 

One could hope that if the private key 
$q$ is chosen randomly, it is impossible to
recover $q$ and the message. This
intuition is wrong. As an easy instance of the 
knapsack problem, the initial sequence $w_i$  carries information
and this information is still
present in the ciphertext in a hidden form. 
This makes it possible to break the system.
For instance, in the Merkle-Hellmann scheme, $w_i$ is 
a superincreasing sequence and Shamir has shown that 
it is possible to recover the initial message $m$, even if 
the private key $q$ remains unknown. 

Thus, starting from an easy instance and hiding it with a random
private key is structurally weak. 
Information can leak, whatever the random choice
of the private key. 

Another potential weakness of knapsack cryptosystems is the 
possibility of low density attacks. 

Usually the numbers $(v_i)_{i \leq s}$ used as the public key are large numbers and
the density $d=s/max \log_2(v_i)$ is low. In this case, 
the elements $(\epsilon_i)$ of the translated lattice $L$ defined by the equation 
$\sum \epsilon_i v_i=C(m)$ are expected to be large,
and the plaintext message $m$ sent by 
Bob to Alice is expected to be the smallest element in $L$.
Besides this heuristic argument, 
this circle of ideas yields
a provable reduction of the knapsack problem
to the closest vector problem CVP  
( CVP consists in finding the closest point  to a fixed point $P$
in a lattice). % When the lattice is sufficiently 
% general, the algorithms which approximate 
% well enough the solution
% of CVP return indeed the solution to CVP. 
In particular, using polynomial time
algorithms to approximate CVP \cite{babaiCVP86}, 
the knapsack problem is solvable in polynomial time when
the density is low enough and  the knapsack is sufficiently
general : most knapsacks of density roughly
less than $2/s$ are solvable in polynomial time 
\cite{nguyenStern01:twoFaces}
.

When the density is low but not less than $2/s$, there is no known
polynomial time algorithm to solve knapsack problems. However, 
one can still reduce knapsack problems
to CVP. The embedding method reduces CVP to the shortest vector
problem SVP 
 with high probability when the density $d$ of the
knapsack is low enough, explicitly when $d\leq 0.9408...$
( SVP consists in finding the shortest vector
in a lattice). 
Although CVP is NP hard and SVP is
NP-hard under randomized reductions \cite{nguyenStern01:twoFaces}, there are algorithms 
which solve efficiently CVP and SVP in low dimension, notably  LLL
based-algorithms. In practical terms, 
a knapsack cryptosystem should have dimension $s$ at least 300 
to avoid such attacks. 

%the
%existence of the LLL %and BKZ 
%algorithm  and its variations
%suggests that the minimum acceptable level of security
%is in dimension $s\geq 300$
%(\cite{}). %sometimes more depending on the encryption scheme. 

\subsection*{Aim of the article}
\label{sec:content-paper}

Summing up, Alice constructs a cryptosystem starting from an instance
$(w_i)_{i \leq s}$ and hides it with a private key $q$. The public key $v_i=v_i(q,w_i)$ is a
function of $q$ and $w_i$.  
The above analysis shows that a knapsack
cryptosystem is potentially weak if one 
starts with an easy instance $(w_i)_{i \leq s}$. To construct a 
robust cryptosystem, one should start with a 
hard instance $(w_i)_{i \leq s}$, ie the $w_i$'s should have no
structure (chosen
randomly). The dimension $s$ should be at least $300$. 
Under these conditions, 
breaking the cryptosystem should be as difficult 
as recovering the private key $q$ since the existence of the private
key is the only reason which makes the message received by Alice 
decipherable. In particular, the difficulty to find the private 
key is expected to be a 
measure of the security of the system.

%In particular, if recovering the key $q$
%is a $NP$-complete problem, the cryptosystem should be a NP-complete
%cryptosystem. 

The goal of this paper is to construct such cryptosystems which 
start with a random instance $(w_i)_{i \leq s}$ in high dimension  $s$
and such that finding the private key is as difficult as factorising a product of two
primes. 

Unlike the other knapsack cryptosystems, our construction does not
include modular multiplications.

\subsection*{Differences and similarities between the three cryptosystems}
\label{sec:diff-simil-betw}

The first of our three systems is the most natural. It is a fast
system, both for encryption and decryption. The
drawback is the size of the public key which goes from 0.1MB to 4.9MB
depending on the level of security considered. 

The size of the public key is subject to debate. Some authors
want a short key. Other authors (see
\cite{goldreichGH97:cryptoParLattice}) 
think that the concept of a small
key should be questioned, and that, in view of the transmission rates on the 
Internet today, it is preferable to have a fast and
secure system than a system with a small public key. 

The sizes of the keys considered in the first system 
are large. Though they could be compatible with
the transmission rates on the internet or the size of the memory of
modern computers, it is nevertheless 
desirable to shorten the keys. We thus construct a second system 
based on the same ideas with a shorter key. The size of the key 
starts from 0.03MB  for a reasonably secure system ( corresponding to 
a knapsack problem with $s=500$ elements),
and is around 0.1MB in dimension $s=1000$. 

Our third cryptosystem is a hybrid between the two first
cryptosystems. The key is not much longer than in the second
cryptosystem, but the private key has been hidden more carefully 
and the system is more secure.

Our three cryptosystems have in common the same 
underlying one-way function based on the following 
remark: it is fast to 
produce divisions $n_i=qx_i+r_i$ with small rests $r_i<<q$ (choose
$q,x_i,r_i$ and compute $n_i$)  
but it takes more time to recover the divisions once 
the numbers $n_i$ are given. 
For instance, if there is one number $n$ 
and we look for the smallest rest $r=0$ in a division $n=qx+r$, it means
that we try to find a factorisation of $n$. The security of the 
RSA system relies on the difficulty to factorize a  product of two
primes $n=qx$. Thus our one way function 
can be seen as a generalisation of the one way function used in the RSA
system.  Section \ref{sec:one-way-function} explains this one-way function with
more details.

%To be precise, fix $N>>0$ and consider 
%a euclidian division $n=xq+r$ with a small rest in the sense that
%$Nr>q$. If $q$ is the only non trivial divisor which yields a division
%with a small rest $r$, then the data $(n,N)$
%and $(x,q,r,N)$ are equivalent. However, it is faster to compute $n$
%from $(x,q,r)$ than to compute $(x,q,r)$ from $n$. Heuristically, it is 
%possible to break the system if it is fast enough to find a divisor $q$ which yields 
%a small rest. When $N$ tends to infinity, this heuristic suggests that
%breaking the system is as difficult as finding a non trivial divisor
%of an integer $n$. This heuristic is correct and we prove a security 
%result in this direction (see theorem \ref{} below). 

\section*{The results} 

We provide complexity results, experimental results, and security
results for the cryptosystems. 

\subsection*{Complexity results}
\label{sec:quantitative-results}

There are various possible choices for the parameters.
There are two base parameters $s,p$, with $s=o(p)$ 
and the other parameters
depend on $s$ and $p$.   
The complexity results for the first system are as follows,
where $\epsilon$ is an arbitrarily small positive number.

\begin{thm} \ \\
  Size of the public key $x_s$: $O(s^2\log_2(p))$\\
Size of the private key $\epsilon,q_i,\sigma,\tau$ : $O(s^2\log_2(p))$ \\
Encryption time: $O(s^2\log_2(p))$\\
Decryption time: $O(s^2\log_2(p))^{1+\epsilon}$\\
Creation time of the public key: $O(s^3\log^2(p)^{1+\epsilon})$\\
Density of the knapsack associated with $x_s$: $1/\log_2(p)$.  
\end{thm}

The complexity results for the second system are the following: 

\begin{thm}
  Size of the public key $x_1$: $O(s^2+s\log_2(p))$\\
Size of the private key : $O(s^2+ s \log_2(p))$ \\
Encryption time: $O(s^2+s\log_2(p))$\\
Decryption time: $O(s^2+\log_2(p)^{1+\epsilon})$\\
Time to create the public key: $O(s^2+\log^2(p)^{1+\epsilon})$\\
Density of the knapsack associated with $x_s$: $\frac{1}{1+\frac{2}{s}+\frac{2\log_2(p)}{s}})$.  
\end{thm}

For the parameters chosen as in variant $2$, we have:
\begin{thm}
  Size of the public key $x_1$: $O(s^2\log_2(p))$\\
Size of the private key : $O(s^2+ s \log_2(p))$ \\
Encryption time: $O(s^2+s\log_2(p))$\\
Decryption time: $O(s^2+\log_2(p)^{1+\epsilon})$\\
Time needed to create the public key: $O(s^2+s\log^2(p))$\\
Density of the knapsack associated with $x_s$: $\frac{1}{2+\frac{2}{s}+\frac{\log_2(p)}{s}})$.  
\end{thm}

By construction, the third system is a hybrid mixing the first and
second system. For brevity, we have not included its complexity
results which can be computed as for the previous two systems.

\subsection*{Experimental results for the first system}

\label{sec:experimental-results}
We report experiments to show that encryption/decryption 
time is acceptable in high dimension. 
The processor used is an Intel Xeon at 2GHz. The programs have been 
written with the software Maple (slow high level language manipulating
nativly arbitrarily large integers).

\textbf{Encryption time in seconds}
$\left| \begin{array}{c|c|c|c|c|c}
\hline
s\backslash p&10^{6}&10^{9}&10^{12}&10^{15}&10^{18}\\ 
200 &    0.002 &    0.001 &    0.001 &    0.001 &    0.001 \\ 
400 &    0.001 &    0.001 &    0.001 &    0.002 &    0.002 \\ 
600 &    0.001 &    0.002 &    0.002 &    0.002 &    0.144 \\ 
800 &    0.003 &    0.002 &    0.003 &    0.004 &    0.261 \\ 
 \hline \end{array}\right|$

\textbf{Decryption time in seconds}
$\left| \begin{array}{c|c|c|c|c|c}
\hline
s\backslash p&10^{6}&10^{9}&10^{12}&10^{15}&10^{18}\\ 
200 &    0.150 &    0.152 &    0.166 &    0.178 &    0.209 \\ 
400 &    0.480 &    0.481 &    0.587 &    0.872 &    0.872 \\ 
600 &    1.019 &    1.025 &    1.182 &    2.343 &    2.099 \\ 
800 &    1.597 &    1.602 &    1.809 &    3.813 &    3.314 \\ 
 \hline \end{array}\right|$

\textbf{Time for generating the key in seconds}
$\left| \begin{array}{c|c|c|c|c|c}
\hline
s\backslash p&10^{6}&10^{9}&10^{12}&10^{15}&10^{18}\\ 
200 &    0.543 &    0.602 &    0.713 &    0.850 &    0.965 \\ 
400 &    3.121 &    3.707 &    4.984 &    9.933 &   11.155 \\ 
600 &   12.127 &   14.164 &   18.500 &   46.012 &   52.045 \\ 
800 &   25.940 &   31.376 &   37.769 &  113.364 &  118.746 \\ 
 \hline \end{array}\right|$

\textbf{Size of the key in MegaBytes}
$\left| \begin{array}{c|c|c|c|c|c}
\hline
s\backslash p&10^{6}&10^{9}&10^{12}&10^{15}&10^{18}\\ 
200 &    0.107 &    0.157 &    0.207 &    0.257 &    0.307 \\ 
400 &    0.430 &    0.628 &    0.828 &    1.027 &    1.226 \\ 
600 &    0.966 &    1.413 &    1.864 &    2.312 &    2.760 \\ 
800 &    1.720 &    2.514 &    3.312 &    4.111 &    4.908 \\ 
 \hline \end{array}\right|$

%\textbf{Erreurs}
%\input{/users/evain/perso/maths/articles/NPcomplet/erreur}

\subsection*{Experimental results for the second system.}
\label{sec:tabl-second-syst}

\textbf{Encryption time in seconds}
$\left| \begin{array}{c|c|c|c|c|c}
\hline
s\backslash p&10^{6}&10^{9}&10^{12}&10^{15}&10^{18}\\ 
500 &    0.002 &    0.001 &    0.001 &    0.000 &    0.001 \\ 
800 &    0.001 &    0.002 &    0.002 &    0.001 &    0.001 \\ 
1100 &    0.001 &    0.001 &    0.001 &    0.008 &    0.002 \\ 
1400 &    0.002 &    0.001 &    0.002 &    0.002 &    0.001 \\ 
1700 &    0.001 &    0.002 &    0.002 &    0.002 &    0.002 \\ 
2000 &    0.002 &    0.002 &    0.003 &    0.003 &    0.002 \\ 
 \hline \end{array}\right|$

\textbf{Decryption time in seconds}
$\left| \begin{array}{c|c|c|c|c|c}
\hline
s\backslash p&10^{6}&10^{9}&10^{12}&10^{15}&10^{18}\\ 
500 &    0.003 &    0.007 &    0.001 &    0.002 &    0.002 \\ 
800 &    0.003 &    0.003 &    0.003 &    0.003 &    0.003 \\ 
1100 &    0.004 &    0.003 &    0.003 &    0.003 &    0.003 \\ 
1400 &    0.005 &    0.004 &    0.005 &    0.005 &    0.004 \\ 
1700 &    0.014 &    0.005 &    0.005 &    0.006 &    0.006 \\ 
2000 &    0.015 &    0.006 &    0.006 &    0.007 &    0.006 \\ 
 \hline \end{array}\right|$

\textbf{Time for generating the key in seconds}
$\left| \begin{array}{c|c|c|c|c|c}
\hline
s\backslash p&10^{6}&10^{9}&10^{12}&10^{15}&10^{18}\\ 
500 &    0.056 &    0.056 &    0.057 &    0.058 &    0.057 \\ 
800 &    0.091 &    0.092 &    0.094 &    0.094 &    0.094 \\ 
1100 &    0.129 &    0.127 &    0.133 &    0.127 &    0.125 \\ 
1400 &    0.166 &    0.168 &    0.165 &    0.169 &    0.169 \\ 
1700 &    0.199 &    0.198 &    0.205 &    0.203 &    0.210 \\ 
2000 &    0.239 &    0.237 &    0.244 &    0.245 &    0.254 \\ 
 \hline \end{array}\right|$

\textbf{Size of the key in MegaBytes}
$\left| \begin{array}{c|c|c|c|c|c}
\hline
s\backslash p&10^{6}&10^{9}&10^{12}&10^{15}&10^{18}\\ 
500 &    0.034 &    0.035 &    0.036 &    0.037 &    0.039 \\ 
800 &    0.084 &    0.086 &    0.088 &    0.090 &    0.092 \\ 
1100 &    0.157 &    0.159 &    0.162 &    0.165 &    0.168 \\ 
1400 &    0.252 &    0.255 &    0.259 &    0.262 &    0.266 \\ 
1700 &    0.370 &    0.374 &    0.378 &    0.382 &    0.387 \\ 
2000 &    0.510 &    0.515 &    0.520 &    0.525 &    0.530 \\ 
 \hline \end{array}\right|$

%\textbf{Erreur}
%\input{/users/evain/perso/maths/articles/NPcomplet/erreurCleCourte}

\subsection*{Security  results}
\label{sec:qualitative-results}

We now come to the security analysis of the cryptosystems. 
Among the three cryptosystems described, 
it is easier to attack the second cryptosystem (shortest key, built 
to be fast, no special care to hide the private key).  
Thus we concentrate our analysis for this second system. 

First, we remark on the above formulas that 
the density can be as close to 1 as possible with
a suitable choice of the parameters. Thus the parameters can be chosen
to avoid low density attacks. 

We consider both exact cryptanalyse and heuristic attacks. 

We show that finding the private key $q$ is as difficult as
factorising a number $n$ which is a product of two primes:
  if it is possible to find the private key $q$ in polynomial time, then
  $\forall \eta >0$, it is possible to factorise $n=pq$ in
  polynomial time with a probability of success at least
  $1-\eta$ (theorem \ref{trouverCle=factoriserUnNombre}). 

In fact, our result is a little more precise. 
The private key $q$ is an integer with suitable properties.
One could use a ``pseudo-key'' $q'$, ie. an integer with the same
properties as $q$,
%(possibly but not necessarily different from $q$) 
to cryptanalyse
the system. Our result says that finding a pseudo-key $q'$ with
the help some extra-information is as difficult  as factorising 
a product of primes (ie. there is a
polynomial probabilistic reduction as above). 
Moreover, the system is more secure if 
$q$ is the only integer with the required properties. 
We give evidences in section \ref{sec:finding-private-key}
that one can construct with 
high probability a cryptosystem with $q$ as the only pseudo-key.

%Thus one could maybe hope that one could find a pseudo-key $q'$
%with the same properties as $q$ but different from $q$. There could be many
%possible values of $q'$, making the system more easy to attack. 
%However, we prove by an heuristic argument that 
%the probability to find a pseudo-key $q'$ with random attempts is  
%is intractable in practice (\ref{} and \ref{}). 
%We expect in general that for sensible choices of the parameters, 
%$q$ is unique element wich has the necessary properties to decypher. 

The above results express that it is difficult to find a 
pseudo-key. But the
cryptosystem could still be attacked by heuristic attacks. Since most
heuristic attacks rely on the LLL-algorithm and its improvements,
we consider the standard attack
relying on the LLL-algorithm and the embedding method.

NP-completness and many experiments lead to the conclusion 
that the knapsack problem is not solvable for a random instance
$x_0=(v_1,\dots,v_s)$ in high dimension $s$.
The public key is not a random instance $x_0$ but 
a slight deformation $x_1$ of $x_0$. A weakness appears if 
the heuristic attacks perform better when the random $x_0$ is replaced by $x_1$. 

Our result  (theorem \ref{thm:comparaisonCleGeneriqueEtSpeciale}) 
says in substance that, if $x_0$ is
very general, replacing $x_0$ by 
a suitable $x_1$ is not dangerous : both the number of steps to perform the
algorithm and the probability of success are unchanged. 
In other terms, with respect to LLL-attacks, the system is as secure if the message is 
cyphered with $x_0$ or with a suitable $x_1$. 
% Basically, the reason is 
% that $x_1$ can be chosen close to $x_0$ in the underlying projective space, 
% and that the LLL algorithm acts on the projective space in a 
% somehow continuous manner. 

%\subsection*{Final remarks}
%I am not an expert in cryptography, thus it could well happen
%that the vocabulary I employed or the citations of previous work could
%be improved. I will appreciate any comment.

\subsubsection*{Acknowledgments}
\label{sec:acknowledgments}
Nice surveys on knapsack cryptosystems made the subject
accessible to me. I am in particular grateful to the authors of 
\cite{nguyenStern01:twoFaces},
\cite{odlyzko89:cryptoSurvey} and 
\cite{brickellOdlyzko92:cryptoSurvey}.

\section{First system}
\label{sec:system}

\subsection{Description of the system}
\label{sec:description-system}

We denote by $M_{p\x q}(A)$ the set of $p\x q$ matrices with
coefficients in the set $A$. 

\begin{listecompacte}
\item \textbf{List of parameters:}$M,s\in\NN$, $\epsilon\in M_{s\x
    s}(\NN)$, $p_1,\dots,p_s,q_1,\dots,q_s\in\NN$, $x_0\in M_{1\x s}(\NN)$, 
\item \textbf{Message to be transmitted:} a column vector $m\in
  \{0,1,\dots,M-1\}^s=M_{s\times 1}(\{0,\dots,M-1\})$.
\item \textbf{Private key:}
  \begin{listecompacte}\item 
    An invertible matrix $\epsilon\in M_{s\x s}(\NN)$
    with rows $\epsilon_1,\dots,\epsilon_{s}$. We let
    $||\epsilon_i||_1=\sum_{j=1}^{j=s} \epsilon_{ij}$ the norm of the
    $i^{th}$ row.
  \item A $s$-tuple of
    positive rational numbers $\lambda_i=\frac{p_i}{q_i}$,$i=1 ,\dots ,s$ such that
    $(M-1)\lambda_i||\epsilon_i||_1 < 1$.
  \end{listecompacte}
\item \textbf{Recursive Construction:} Choose a random row vector $x_0\in \NN^s$.
Define the row vector $x_i$,
  $i=1\dots s$ by $x_i=q_{i}x_{i-1}+p_{i}\epsilon_{i}$.
\item \textbf{Public key:} $x_s$
\item \textbf{Cyphered message}: $x_sm\in \NN$. 
\end{listecompacte}

\begin{nt}
  We denote by $C$ the cyphering function $\{0,1,\dots,M-1\}^s \fd
  \NN$, $m\mapsto N_s=x_s.m$
\end{nt}

\begin{prop}
  The function $C$ is injective. 
\end{prop}
  It suffices to explain how to decypher to prove the proposition.
  We define $N_i$, $0 \leq i\leq s$ and $O_i$, $1\leq i \leq s$ by
  decreasing induction:
  \begin{listecompacte}
  \item $N_s=C(m)=x_sm$
  \item $N_{i-1}=[\frac {N_i}{q_{i}}]$, where $[.]$ denotes the
    integer part
  \item $O_i=(N_i-q_iN_{i-1})/p_i$.
  \item Let $N\in M_{s+1\x 1}(\NN)$ be the column vector with entries
    $N_0,\dots,N_s$
  \item Let $O\in M_{s\x 1}(\QQ)$ be the column vector with entries
    $O_1,\dots, O_s$.
  \item Let $X\in M_{s+1\x s}(\NN)$ be the matrix with rows
    $x_0,\dots,x_s$.
  \end{listecompacte}

\begin{prop} \label{preuveDechiffrement}
 The message $m$ verifies $Xm=N$, $\epsilon m=O$. In particular,
  the coefficients of $O$ are integers. 
\end{prop}
\begin{dem}
  %The vector space generated by the rows of $X$ is the same as the
  %vector space generated by $(\epsilon_1,\dots,\epsilon_{s-1},x_1)$,
  %thus $X$ is invertible. 
We prove that $x_i m=N_i$ by
  decreasing induction on $i$. The case $i=s$ is true by definition.
  If $x_im=N_i$,
  then $(x_{i-1}+\lambda_{i}\epsilon_{i})m=N_i/q_{i}$. Since
  $x_{i-1}m\in \NN$ and $0<\lambda_{i}\epsilon_{i}m\leq \lambda_i
  ||\epsilon _i||_1 (M-1)<1$ by
  hypothesis, we obtain  $x_{i-1}m=[N_i/q_{i}]=N_{i-1}$, as expected.
 Thus $\epsilon_i m=
 (x_i-(q_{i}x_{i-1}))m/p_{i}=(N_i-q_iN_{i-1})/p_i=O_i$.
\end{dem}

\begin{coro} \label{dechiffrementSysteme1}
  To decypher the message,
  \begin{listecompacte}
  \item  Compute $N_{s-1},\dots,N_1$ with the formula $N_{i-1}=[\frac {N_i}{q_{i}}]$.
 \item Compute $O_i=(N_i-q_iN_{i-1})/p_i$.
\item Solve the system  $\epsilon m=O$.
  \end{listecompacte}
\end{coro}

\subsection{Analysis of the system}
\label{sec:analysis-system}

\subsubsection*{The underlying one way function}
\label{sec:one-way-function}

We make a quick analysis of the system.

The couple $(q_s$, $\epsilon_s)$ in the private key satisfies
$x_s=q_s x_{s-1}+p_s \epsilon_s$ with $q_s> p_s ||\epsilon_s||_1(M-1)$. 
Componentwise, $p_s\epsilon_{si}$ is the rest of the division of
$x_{si}$ by $q_s$. These rests are small. The rest of the division
of $x_{si}$ by $q_s$ is at most $q_s$, and the sum of the rests $p_s
\epsilon_{si}$ for $1\leq i\leq s$  is at most $sq_s$ in general. 
In the present situation, 
the sum $\sum_i p_s \epsilon_{si}=p_s ||\epsilon_s||_1$ 
of all the rests is 
at most $\frac{q_s}{M-1}$.

In other
words, an eavesdropper who tries to break the system
looks for an integer $q_s$ 
such that the rests of the divisions of the $x_{si}$ by $q_s$ 
are unusualy small: the sum of the $s$ rests is at most
$\frac{q_s}{M-1} $.

There is hopefully a one way function here.
It is easy to construct a couple of integers $(x,q)$ such that the
rest of the division of $x$ by $q$ is small. But once $x$ is given, it
is not easy to find back an integer $q$ such that the rest of the 
division of $x$ by $q$ is small. 

For instance, to obtain a 
rest which is at most $\frac{1}{10^n}$ of the divisor $q$, choose any $y,q\in \NN$,
$0\leq \epsilon \leq q/10^n$ and put $x=qy+\epsilon$. 
As a function of $q$, the number of operations to compute $x$ is
$O(log_2(q))$. 
If $x$ is given and Eve knows
that there is a $q$ satisfying $x=qy+\epsilon$, $10^n\epsilon<q$,
trying successivly all possible divisors $1,\dots,q$
requires $O(q)$ operations. 

Thus, in the absence of a quick algorithm to find $q$, 
there is a gain of an exponential factor 
here. In our choice of parameters, the numbers $q_i$ will be
large to make the most of this advantage.

\subsubsection*{Construction of the matrix $\epsilon$}
\label{sec:constr-matr-epsil}

The matrix $\epsilon$ of the private key 
should be quickly invertible, for instance triangular, to
facilitate decryption (see corollary \ref{dechiffrementSysteme1}). 
But a triangular matrix $\epsilon$, or any matrix with 
a lot of null coefficients, would be a bad choice.
Indeed, if $\epsilon$ is sparse, there are 
two components $c,c'$ of 
$x_s=q_sx_{s-1}+p_s\epsilon_s=(....,c,....,c',....)$ 
whose gcd is a multiple of $q_s$, or $q_s$ itself.
After several attempts, the
eavesdropper could find $q_s$. 

The same problem occurs if the components of $\epsilon_s$ 
are too small or well localised by a law of repartition.
If $x_s=(\dots,c,\dots,c',...)$, there is a natural
attempt to find $q_s$:   test for the gcd of $
(c-\epsilon',c'-\epsilon'')$ for several values of
$\epsilon',\epsilon''$. 

Summing up, the matrix $\epsilon$ should satisfy the two following 
conditions:
\begin{listecompacte}
 \item its coefficients are difficult tolocalize,
\item solving $\epsilon m=O$ is fast.
\end{listecompacte}
If the coefficients of the matrix $\epsilon$ are chosen
randomly, it takes time to solve $\epsilon m=O$. If we choose a lower triangular
matrix $L$, an upper triangular matrix $U$ with random uniform
coefficients, and choose $\epsilon=LU$, then it is easy to solve the 
system  but the
coefficients of $\epsilon$ are not random uniform and this non
uniformity could be used to cryptanalyse the system as explained
above.  

Thus there is a compromise to find between the amount of time required
to compute and invert $\epsilon$ and the uniformity 
in the coefficients of $\epsilon$. 
Our approach to find the compromise is to consider an upper triangular
matrix $U$ with random coefficients and to deform it using elementary
operations (proposition \ref{description de la matrice}).

Let $L,N \in M_{s\x s}(\NN)$ be the lower triangular matrices defined 
by $L_{ii}=N_{ii}=1$, $L_{i,1}=1$, $N_{n,i}=1$ and all other
coefficients equal to zero. If $\sigma$ is a permutation of $\{1,\dots,s\}$, 
we denote by $M_{\sigma}$ the permutation matrix defined by
$M_{i,\sigma(i)}=1$ and $M_{ij}=0$ otherwise. 

\begin{prop}\label{description de la matrice}
  Let $U\in M_{s\x s}(\NN)$ be an upper invertible triangular matrix with
  coefficients $u_{ij} $, $i\leq j$ chosen randomly in $\{1,\dots,x\}$
and $\sigma,\tau$ be permutations of $\{1,\dots,s\}$. 
Then every entry $e$  of the matrix $\epsilon(s,x)=M_{\sigma}LUNM_{\tau}$ 
verifies $0\leq e \leq 4x$. In particular, the norm of the lines
$\epsilon_i$ satisfy $||\epsilon_i||_1\leq 4sx$. 
\end{prop}
\begin{dem}
  The action of the permutations $\sigma,\tau$ permute the coefficients of
  $LUN$ so one can suppose $\sigma=\tau=Identity$. An entry in $U$
  is in $\{0,\dots,x\}$. The left
  multiplication with $L$ replaces a line $L_i,i>1$ with $L_i+L_1$. The
  right multiplication with $N$ replaces a column $C_i,i<s$ with
  $C_i+C_s$. Thus an entry of $LUN$ is in $\{0,\dots,4x\}$.   
\end{dem}

\subsection{Suggested choice for the parameters}
\label{sec:choice-parameters-1}

In this section, 
suggestions for our list of 
parameters $M,s\in\NN$, $\epsilon\in M_{s\x
    s}(\NN)$, $p_1,\dots,p_s,q_1,\dots,q_s\in\NN$, $x_0\in M_{1\x
    s}(\NN)$ are given. We fix 
  two integers $s,p$ as based parameters.
The other parameters are constant or functions of $s$ and
  $p$. 

The level of security depends on
  the size of $s$ and $p$. To give an idea of the size of the numbers
  involved, $s>300$ and
  $p>10^6$ are sensible choices.  

\subsubsection*{Suggested choice for the parameters as constants or 
functions of $s,p$:}
\begin{listecompacte}
\item $M=2$
\item $\epsilon= \epsilon(s,[p/4s])$ is the random matrix considered
  in proposition  \ref{description de la matrice}.
\item $p_i=1$, $q_i$ chosen randomly in $[p+1,2p]$ (uniform law)
\item $x_0$ has entries chosen randomly in $[0,2^s]$ (uniform law)
\end{listecompacte}

\subsubsection*{Comments on the choices.}
The choice $M=2$ is to make the system as simple as possible. 
Moreover, Shamir has shown that compact knapsack cryptosystems
(ie. those with messages in $\{0,\dots,M-1\}^s$ and small $M$) tend to
be more secure \cite{shamir79:compactKnapsacks}. 

The reason for the choice of the matrix $\epsilon$ 
has been given before proposition
\ref{description de la matrice} 
(compromise between randomness and inversibility). 
Note that the 
required condition $(M-1)||\epsilon_i|| \lambda_i <1$ is satisfied
by proposition \ref{description de la matrice}.

As to the choice of $\lambda_i=\frac{p_i}{q_i}$, we have explained
that $q_i$ is large to make the most of the one way function.
Looking at the recursive definition of $x_i$, 
it appears that the $x_i$'s are large when $p_i$ is large.
Thus we take $p_i=1$ to limit the size of the key.  

The entries of the initial vector $x_0$ are chosen randomly in
$[0,2^s]$ so that the density of the 
knapsack cryptosystem associated to $x_0$ is expected close to one. 
If the density is lower, there could be a low density attack on 
$x_0$, and maybe an attack on $x_s$ as $x_s$ is a
modification of $x_0$. On the other hand, it is not clear that 
a higher density is dangerous. It could even be a better choice. 
Experiments are needed to decide. Thus we propose a variant of higher
density:

\subsubsection*{Variant for the choice of parameters}
\begin{listecompacte}
\item  $x_0$ has entries chosen randomly in $\{0, \dots, s^5\}$. 
\item All other parameters are chosen as before. 
\end{listecompacte}

\subsection{Complexity results}
\label{sec:complexity-results}

The complexity of the cryptosystem is described in the following
theorem, using the first variant for the choice of parameters
(ie. $x_0$ has entries in $\{0,\dots,2^s\}$).

We denote by $size(A)$ the number of bits needed to
  store an element $A$ and by $time(A)$ the number of elementary operations
  needed to compute $A$. Recall that, for all $\epsilon>0$, 
computing a multiplication of two integers $p$ and $q$ takes 
$time(pq)=O(size(p)+ size(q))^{1+\epsilon})$ elementary
operations \cite{knuth69:algoSemiNumeriques}. Moreover, the complexity
of a division is the same as the complexity of a multiplication.

\begin{thm} Suppose that $s=o(p)$. Then: \\
  Size of the public key $x_s$: $O(s^2\log_2(p))$\\
Size of the private key $\epsilon,q_i,\sigma,\tau$ : $O(s^2\log_2(p))$ \\
Encryption time: $O(s^2\log_2(p))$\\
Decryption time: $O(s^2\log_2(p))^{1+\epsilon}$\\
Creation time of the public key: $O(s^3\log^2(p)^{1+\epsilon})$\\
Density of the knapsack associated with $x_s$: $1/\log_2(p)$.  
\end{thm}
\begin{dem} 
  \begin{listecompacte}
  \item $||\epsilon_i||_\infty \leq p$
  \item $size(||\epsilon_i||_\infty)=O(\log_2(p))$
  \item $size(\epsilon_i)\leq s\ size(||\epsilon_i||_\infty)=O(s\log_2(p))$ 
  \item  $size(\epsilon)=\sum_i size(\epsilon_i)=O(s^2\log_2(p))$
  \item $size(q_1,\dots,q_s)=O(s\log_2(p))$
  \item $size(\sigma)=size(\tau)=time(\sigma)=time(\tau)= O(s \log_2(s))$
  \item $\mathbf{size(private\ key)}=size(\epsilon,q_1,\dots,q_s,\sigma,\tau)= O(s^2\log_2(p))$
\item $||x_i=q_ix_{i-1}+\epsilon_i||_\infty\leq |q_i|
  ||x_{i-1}||_\infty+||\epsilon_i||_\infty\leq 2p ||x_{i-1}||_\infty +
  p$ thus $||x_i||_\infty \leq 3^ip^i ||x_0||_{\infty}$. 
\item  $size(||x_i||_\infty)=O(i \log_2(p)+size(||x_0||_\infty))=O(i\log_2(p)+s)$
\item $size(x_i)\leq s\;size(||x_i||_\infty)= O(is\log_2(p)+s^2)$
\item $\mathbf{size(public\ key)}=size(x_s)=O(s^2 \log_2(p))$
\item $\mathbf{encryption \ time}=size(public\ key)=O(s^2 \log_2(p))$
\item $time(x_i)=O(size(q_i)^{1+\epsilon}+size(x_{i-1})^{1+\epsilon}+size
(\epsilon_i))=O( size(x_{i-1})^{1+\epsilon}
)=O((is\log_2(p)+s^2)^{1+\epsilon})\leq O((s^2 log_2(p)^{1+\epsilon}))$ 
\item $\mathbf{time(public\ key)}=\sum time(x_i)=O((s^3\log_2(p))^{1+\epsilon})$ 
\item $time(N_i=[N_{i+1}/q_i])= O(size(q_i)^{1+\epsilon}+size(N_{i+1})^{1+\epsilon})
= O(\log_2(p)^{1+\epsilon}+size(x_{i+1}m)^{1+\epsilon})\leq
O(\log_2(p)^{1+\epsilon}+size(s||x_{i+1}||_\infty)^{1+\epsilon})=O(i\log_2(p)+s)^{1+\epsilon}\leq
O((s \log_2(p))^{1+\epsilon})$
\item $time(N_{0},\dots,N_s)=O(\log_2(p) s^2)^{1+\epsilon}$. 
\item $time(O_i=(N_i-q_iN_{i-1}))=O(time(N_i))$ 
\item $time(N_0,\dots,N_{s},O_1,\dots, O_s)=time(N_0,\dots,N_s)=O(\log_2(p) s^2)^{1+\epsilon}$
\end{listecompacte}
To solve the linear
$\epsilon m=O$ with $\epsilon=M_{\sigma}LUNM_\tau$. 
we first suppose that $\epsilon
=U$ (ie. $M_\sigma=L=N=M_\tau=Id$). The entries $e$ in $\epsilon $ and $O$ satisfy
$size(e)=O(\log_2(p))$. 
Since $\epsilon=U$ is triangular, 
solving the system takes a time $\tau=O(s^2\log_2(p))^{1+\epsilon}$.
We have
$time(decryption)=time(N_1,\dots,N_s,O_1,\dots,O_s,solving(\epsilon.m=O))$, thus
 the
decryption 
takes $O(s^2\log_2(p))^{1+\epsilon}$ operations. Since inverting
$M_{\sigma},L,N,M_\tau$ require $O(s^2)$ operations, replacing
$\epsilon=U$ by $\epsilon=M_{\sigma}LUNM_\tau$
does not change the complexity. 
\end{dem}

% \begin{rem}
%   The computation of the public
%   key $x_s$ requires the intermediate computations of
%   $x_1,\dots,x_{s-1}$.
%   Though this computation is done only once, it could take too long on very small
%   systems. A public listing could be used to bypass this problem. For
%   instance, Alice could choose a key
%   $x_s,q_1,\dots,q_s,\epsilon,\sigma,\tau$ in a public
%   listing. Then she would choose $q_{s+1}$ and $\epsilon_{s+1}$ secretly,
%   and would take $x_{s+1}=q_{s+1}x_s+\epsilon_{s+1}$ for the public
%   key. 
% \end{rem}

\begin{rem}
  \begin{listecompacte}
\item These theoretical results are consistent with the experimental
  results of the introduction. 
%\item 
%   It is possible to improve some estimations of the theorem. For
%   instance $size(\epsilon_i)=O(s\log_2(\frac{p}{s}))$. But since
%   in practice $s=o(p)$, there is no difference between
%   $O(s\log_2(\frac{p}{s}))$ and $O(s\log_2({p})$.
\end{listecompacte}
\end{rem}

\section{Second system}
\label{sec:second-system}

\subsection{Description of the system}
Since the size of the key is a bit large, we propose a second system
to reduce the size of the key. The implicit one way function is the
same as before. We only change the private key and take a
superincreasing sequence instead of an invertible matrix.

\begin{listecompacte}
\item \textbf{List of parameters:}$M,s\in\NN$, $\epsilon\in \NN^s$, $p_1,q_1
  \in \NN$, $x_0\in M_{1\x s}(\NN)$, a permutation $\sigma$ of $\{1,\dots,s\}$ 
\item \textbf{Message to be transmitted:} a column vector $m\in \{0,1,\dots,M-1\}^s$.
\item \textbf{Private key:}
  \begin{listecompacte}\item A permutation $\sigma$ of $\{1,\dots,s\}$
  \item 
    A row matrix $\epsilon$ $\in  M_{1 \x s}( \NN)$
    such that the sequence
    $\epsilon_{\sigma(1)},\dots,\epsilon_{\sigma(s)}$ is a
    superincreasing sequence.
  \item A
    positive rational number $\lambda_1=\frac{p_1}{q_1}$, such that
    $(M-1)\lambda_1||\epsilon||_1 < 1$.
  \end{listecompacte}
\item \textbf{Construction:} Choose a random row vector $x_0\in \NN^s$.
Define the row vector $x_1$
  by $x_1=q_{1}x_{0}+p_{1}\epsilon$.
\item \textbf{Public key:} $x_1$
\item \textbf{Cyphered message}: $x_1m\in \NN$. 
\end{listecompacte}

\begin{nt}
  We denote by $C$ the cyphering function $\{0,1,\dots,M-1\}^s \fd
  \NN$, $m\mapsto C(m)=x_1.m$
\end{nt}

\begin{prop}
  The function $C$ is injective. 
\end{prop}
It suffices to explain how to decypher to prove the proposition. 
We define $N_1,N_0$, and $O$ as follows
 \begin{listecompacte}
 \item $N_1=C(m)=x_1m$ 
 \item 
   $N_{0}=[\frac {N_1}{q_{1}}]$
 \item $O=(N_1-q_1N_{0})/p_1$.
 \item Let $N$ be the column vector with entries
$N_0,N_1$.
\item Let $X$ be the matrix with rows $x_0,x_1$.
\end{listecompacte}

The same proof as for proposition \ref{preuveDechiffrement} shows:
\begin{prop}
 The
  initial message $m$ verifies $Xm=N$, $\epsilon m=O$. 
\end{prop}
Now, since $\epsilon_{\sigma(i)}$ is a superincreasing sequence, the 
map $m \mapsto \epsilon m$ is injective and the formula to decypher $m$
expresses $m_{\sigma(i)}$ by decreasing induction on $i \leq s$. 
\begin{prop}\label{prop:calcul du message}\label{dechiffrement2}
  \begin{listecompacte}\item     
  $m_{\sigma(s)}=1$ if $O\geq \epsilon_{\sigma(s)}$ and
  $m_{\sigma(s)}=0$ otherwise
\item $m_{\sigma(i)}=1$ if $O-\sum_{j>i}
  \epsilon_{\sigma(j)}m_{\sigma(j)}\geq \epsilon_{\sigma(i)}$ and $0$
  otherwise. 
\end{listecompacte}
\end{prop}

\subsection{Suggestion for the choice of the parameters}
\label{sec:sugg-choice-param}
The parameters $s$ and $p$ depend on the required level of security
and the other parameters are constant or functions of $s$ and $p$. 

Variant 1. Choose: 
\begin{listecompacte}
\item $\epsilon_{\sigma(1)}\in [0,p[,\epsilon_{\sigma(2)}\in [p,2p[,\dots, \epsilon_{\sigma(s)}\in
  [(2^{s-1}-1)p,2^{s-1}p[$ (uniform law)
\item $x_0$ in $[0,p]$ (uniform law)
\item $p_1=1$, $M=2$
\item $q_1\in [2^sp,2^{s+1}p]$ (uniform law)
\end{listecompacte}
Variant 2. Choose
\begin{listecompacte}
\item $x_0$ in $[0,2^s]$ (uniform law)
\item the other parameters as above.
\end{listecompacte}

\subsection{Complexity results}
As before, we suppose that the parameters $s$ and $p$ satisfy $s=o(p)$.
For the parameters chosen as in variant $1$, we have: 
\begin{thm}
  Size of the public key $x_1$: $O(s^2+s\log_2(p))$\\
Size of the private key : $O(s^2+ s \log_2(p))$ \\
Encryption time: $O(s^2+s\log_2(p))$\\
Decryption time: $O(s^2+\log_2(p)^{1+\epsilon})$\\
Time to create the public key: $O(s^2+\log^2(p)^{1+\epsilon})$\\
Density of the knapsack associated with $x_s$: $\frac{1}{1+\frac{2}{s}+\frac{2\log_2(p)}{s}})$.  
\end{thm}

For the parameters chosen as in variant $2$, we have:
\begin{thm}
  Size of the public key $x_1$: $O(s^2\log_2(p))$\\
Size of the private key : $O(s^2+ s \log_2(p))$ \\
Encryption time: $O(s^2+s\log_2(p))$\\
Decryption time: $O(s^2+\log_2(p)^{1+\epsilon})$\\
Time needed to create the public key: $O(s^2+s\log^2(p))$\\
Density of the knapsack associated with $x_s$: $\frac{1}{2+\frac{2}{s}+\frac{\log_2(p)}{s}})$.  
\end{thm}

For brevity, we include the proof only for variant 1. 
\begin{dem} (for variant 1). 
  \begin{listecompacte}
  \item $||x_1=q_1x_0+\epsilon||_{\infty}\leq
    2^{s+1}p||x_0||_\infty+||\epsilon||_\infty\leq
    2^{s+1}p^2+2^{s-1}p< 2^{s+2}p^2$
  \item $\mathbf{size(public\ key)}=size(x_1)\leq s\; size(||x_1||_\infty)=O(s^2+s \log_2(p))$.
  \item $size(\epsilon)\leq s\log_2(p)+1+2 +\dots +(s-1)
    =O(s^2+s\log_2(p))$.
  \item $size(q_1)=O(s+log_2(p))$
  \item $size(x_0)=O(log_2(p))$
  \item $size(\sigma)=O( s \log_2(s))$
  \item $\mathbf{size(private\ key)}=size(x_0,q_1,\epsilon, \sigma)=O(s^2+s \log_2(p))$.
  \item $\mathbf{encryption\ time}=size(public\ key)=O(s^2+s \log_2(p))$
  \item $size(N_1)\leq log_2(s||x_1||_\infty)=O(s+\log_2(p))$.
  \item $time(N_0)\leq
    O(size(N_1)^{1+\epsilon}+size(q_1)^{1+\epsilon})=O(s^{1+\epsilon}+
    \log_2(p)^{1+\epsilon} )$
  \item $N_0\leq \frac{N_1}{q_1}\leq \frac{2^{s+2}sp^2}{2^sp}=4sp$
  \item $size(N_0)=O(\log_2(s)+\log_2(p))$.
  \item $time(O)=O(size(N_1)+size(q_1)^{1+\epsilon}+size(N_0)^{1+\epsilon})=O(s^{1+\epsilon} +
    log_2(p)^{1+\epsilon})$ since $s\leq p$.
  \item $O-\sum_{j>i}\epsilon_{\sigma(j)}m_{\sigma(j)}\leq \sum_{j\leq i}
\epsilon_{\sigma(j)}\leq p+2p+\dots + 2^{i-1}p<2^i p$.
\item $time(m_{\sigma(i)}$ in proposition \ref{prop:calcul du message}$)=size(O-\sum_{j>i}\epsilon_{\sigma(j)}m_{\sigma(j)})=O( i+\log_2(p))$
\item $time(m)=\sum_{i=1}^s time(m_{\sigma(i)})= O(s \log_2(p)+1+2+\dots + s)=O(s \log_2(p)+s^2)$.
\item $\mathbf{decryption time}=time(N_0,O,m)=O(s^2+\log_2(p)^{1+\epsilon})$.
\item $\mathbf{time(public\
  key)}=time(q_1x_0+\epsilon)=O(time(\epsilon)+time(q_1)+time(x_0)+size(q_1)^{1+\epsilon}+size(x_0)^{1+\epsilon}+size(\epsilon))=
O(size(q_1)^{1+\epsilon}+size(x_0)^{1+\epsilon}+size(\epsilon))$ since
$time(\epsilon)=O(size(\epsilon))$ and similarly for $q_1$ and
$x_0$. Thus $time (public\ key)=O(s^2+\log_2(p)^{1+\epsilon})$ 
\item $density(knapsack)=\frac{s}{\log_2(||x_1||_\infty)}>
  \frac{s}{s+2+2\log_2(p)}=\frac{1}{1+\frac{2}{s}+\frac{2\log_2(p)}{s}}$.
  \end{listecompacte}
\ 
\end{dem}

\section{Third system}
\label{sec:second-variant}

Two cryptosystems have been constructed so far. 
In the second system, the key is shorter than in the first one,
but the system could be less secure because of the superincreasing sequence. 

This section presents a hybrid system, a
compromise between the two previous systems. 
We still use a superincreasing sequence to shorten the key as in the second
system, but the matrix $\epsilon$ has several lines as in the first system to hide 
more carefully the superincreasing
sequence. Hopefully, this is a good compromise
between security and length of the key.

\begin{listecompacte}
\item List of parameters:$M,s\in\NN$, $\epsilon\in M_{2\x s}(\NN)$,
  $p_1,q_1, p_2,q_2 
  \in \NN$, $x_0\in M_{1\x s}(\NN)$, $\sigma$ a permutation of
  $\{1,\dots,s\}$. 
\item \textbf{Message to be transmitted:} a column vector $m\in \{0,1,\dots,M-1\}^s$.
\item \textbf{private key:}
  \begin{listecompacte}\item A permutation $\sigma$ of $\{1,\dots,s\}$
  \item 
    An invertible $2 \x s$ matrix $\epsilon$ with entries in $\NN$
    such that the row $\mu=\epsilon_2-\epsilon_1$ is a superincreasing
    sequence with respect to the permutation $\sigma$, ie. 
    $\mu_{\sigma(1)},\dots,\mu_{\sigma(s)}$ is a superincreasing sequence. 
  \item Two    positive rational numbers $\lambda_i=\frac{p_i}{q_i}$, such that
    $(M-1)\lambda_i||\epsilon_i|| < 1$.
  \end{listecompacte}
\item \textbf{Construction:} Choose a random row vector $x_0\in \NN^s$.
Define the row vectors $x_1$,$x_2$
  by $x_1=q_{1}x_{0}+p_{1}\epsilon_{1}$, $x_2=q_{2}x_{1}+p_{2}\epsilon_{2}$
\item \textbf{Public key:} $x_2$
\item \textbf{Cyphered message}: $N_2=x_2m \in \NN$. 
\end{listecompacte}

To decypher, we define 
$N_1$, $N_0$ and $O_2,O_1$ as before, and $\omega=O_2-O_1$: 
 \begin{listecompacte}
    \item  Compute $N_{1}$ and $N_0$ with the formula $N_{i-1}=[\frac {N_i}{q_{i}}]$.
 \item Compute $O_i=(N_i-q_iN_{i-1})/p_i$.
 \item Compute $\omega=O_2-O_1$
 \item 
Let $N=\left( 
\begin{matrix}
N_0\\N_1\\N_2.
\end{matrix}
\right)\in M_{3\x 1}(\NN)$ and  $X=\left( 
\begin{matrix}
x_0\\x_1\\x_2
\end{matrix}
\right)\in M_{3\x s}(\NN)$
 \end{listecompacte}

The same proof as for proposition \ref{preuveDechiffrement} shows:
\begin{prop}
 The
  initial message $m$ verifies $Xm=N$, $\epsilon m=O$, $\mu m=\omega$. 
\end{prop}
Now, since $\mu$ is a superincreasing sequence, the 
map $m \mapsto \mu m$ is injective and the formula to decypher is 
as in proposition \ref{dechiffrement2}.

\section{Security results}
\label{sec:qualitative-results-1}
In this section, we analyse the security of the second cryptographic
system (section \ref{sec:second-system}).
We concentrate our attention on this system because it
is the easiest system to attack: the key is short and no
special effort has been done to hide the superincreasing sequence.  

We recall the notations. The private key is $q,
\epsilon_1,\dots,\epsilon_n,x_0,\sigma$ where $x_0=(v_1,\dots,v_s)$,
$\epsilon_{\sigma(i)}$ is a superincreasing sequence and $\sum_{i=1}^s \epsilon_i<q$. 
The public key is $x_1=(w_1,\dots,w_s)$ where $w_i=qv_i+\epsilon_i$. 

Obviously, $\epsilon_i=w_i-[\frac{w_i}{q} ]$, and $
\sigma$ is determined by $\epsilon$. In other words, 
 the whole private key is determined by $q$. 
We thus call $q$ the private key.

\subsection{Unicity of the pseudo-key}
\label{sec:finding-private-key}

It is not necessary to find the private key $q$ to
cryptanalyse. 
Any number $q'$ with the same properties as $q$ would do the job. 
We call such a number a pseudo-key. Explicitly, in our context, 
a pseudo-key is
an integer $q'$ such that the numbers $v_i', r_i$ defined by the 
euclidean divisions $w_i=q'v_i'+r_i$ verify
$\sum_{i=1}^s r_i<q'$ and $(r_i)$ is a  superincreasing 
sequence up to permutation.

If there are many pseudo-keys, it is easier to attack the
system.  For instance, in the Merkell-Hellman 
modular knapsack cryptanalysed
by Shamir-Adleman, there were
many pseudo-keys. The strategy of Shamir was 
to find a pseudo-key.

The experiments made on our cryptosystem 
show that usually the pseudo-key is unique. 
We chose  random instances of the parameters
and we count the percentage of cases where the pseudo-key is unique.
Those results suggest that when
$s>200$, which are the cases considered in practice,
the pseudo-key should be unique and equal 
to the private key 
with high probability.

\begin{prop} \label{pseudoClesCompteExperimental}
  Consider the second cryptosystem, variant 2. The results of the
  experiments are as follows.  
  \begin{listecompacte}
  \item $s=5,20<p<35$, the pseudo-key is  unique in $2\ \%$ of the cases.
  \item $s=6,30<p<45$, the pseudo-key is  unique in $46\ \%$ of the cases.
\item $s=7,30<p<45$, the pseudo- key is unique in $79\ \%$ of the cases.
\item $s=8,40<p<55$, the pseudo-key is  unique in $96\ \%$ of the cases.
  \end{listecompacte}
\end{prop}

Besides this computation, 
we want to explain why we expect
a unique pseudo-key when $s$ is large enough.

For a fixed $q'$, the rests $r_i=w_i \mod
q'$ are numbers between $0\dots q'-1$. In the absence of relation
between $w_i$ and $q'$,  these rests are expected to follow a
uniform law of repartition in $\{0,\dots,q'-1\}$. 
Of course the exact law of $r_i=w_i \mod q'$ depend on the law of
$w_i$ (hence of the law of $q,v_i,\epsilon_i$ as
$w_i=qv_i+\epsilon_i$) 
and of the choice of $q'$, but  a uniform law is 
an approximation for the law of $r_i$.

If one accepts this approximation,
the next proposition is an estimation of the probability to 
find a $q$ such that the sum of the rests is bounded by $q$, as
required for a pseudo-key.

\begin{prop}\label{compteDesClesDePetitReste}
  Let $q\geq 2$. Consider the rests $r_1(q),\dots,r_s(q)$ where  $r_i(q)=w_i \ mod\
  q$. Suppose that $r_1(q),\dots,r_s(q)$ follow independant uniform
  laws with values in $\{0,\dots,q-1\}$. 
  The probability $P$ that $\sum_{i=1}^s r_i(q)\leq
  q-1$ satisfies $P\leq (\frac{3}{4} )^{s-1}$
\end{prop}

\begin{lm}
  Let $a_1\geq a_2\geq\dots\geq a_n$ and $p_1\leq p_2 \leq \dots \leq
  p_n$. 
  Then $n\sum_{i=1}^n a_i
  p_i \leq  (\sum_{i=1}^n a_i)(\sum_{i=1}^n p_i)$. 
\end{lm}
\begin{dem} \textit{of the lemma}
  $(\sum_{i=1}^n a_i)(\sum_{i=1}^n p_i)-n \sum_{i=1}^n a_ip_i= \sum_{i=1}^n a_ip_i +\sum
  _{i=1}^{i=n}a_i\sum_{k=1,k\neq i}^{k=n}p_k-\sum_{i=1}^n a_i p_i-(n-1)\sum_{i=1}^n
  a_ip_i= \sum_{i=1}^{n}\sum_{k=1,k\neq
    i}^{k=n}a_i(p_k-p_i)=\sum_{1\leq i<k \leq n}(a_i-a_k)(p_k-p_i)\geq
  0$. 
\end{dem}

\begin{dem} \textit{of proposition \ref{compteDesClesDePetitReste}}
  We have 
  $P(r_i(q)=k)=\frac{1}{q} $ for every $k\in \{0,\dots,q-1\}$.
For $0\leq r\leq q-1$, denote by $P_{q,s,r}$ the probability that $\sum_{i=1}^s
r_i(q)=r$. We show by induction on $s\geq 1$ that
$P_{q,s,0}\leq P_{q,s,1}\dots \leq P_{q,s,q-1}$ and that
$\sum_{r=0}^{r=q-1}P_{q,s,r}\leq (\frac{3}{4} )^{s-1}$. This is obvious
for $s=1$. Note that $P_{q,s,r}=\frac{\sum_{k=0}^rP_{q,s-1,k}}{q}$. In
particular, 
$\sum_{r=0}^{r=q-1}P_{q,s,r}=\frac{qP_{q,s-1,0}+(q-1)P_{q,s-1,1}+\dots
  + P_{q,s-1,q-1}}{q} \leq \frac{q+1}{2} \frac{P_{q,s-1,0}+\dots
  +P_{q,s-1,q-1}}{q} $ by the lemma. Now the induction implies that
the right hand side of the
inequality is bounded by $\frac{q+1}{2q} (\frac{3}{4}
)^{s-2}\leq (\frac{3}{4} )^{s-1}$ for $q\geq 2$.
\end{dem}

%  The above proposition measures the probability to have small
%  rests. The probability to find a pseudo-key $q$ is in fact much smaller than
%  the above estimation since
%  we have not taken into account that the rests must form a
%  superincreasing sequence. 

\begin{prop} \label{nombreDeSuperCroissantes}
Let $s\in \NN$ be a fixed number and $t>>s$. Let $S_{st}$ the number
of superincreasing sequences $r_1,\dots,r_s$ with sum $t$ and
$C_{st}$ the number of sequences with sum $t$. Then 
$\frac{C_{st}}{S_{st}}$ is asymptotically equal to
$\frac{1}{2^{\frac{s(s-1)}{2}}} $
when $t$ tends to infinity.     
\end{prop}
\begin{dem}
  The number of sequences $r_1,\dots,r_s$ with sum $t$ is $t+s-1
  \choose s-1$ and is equivalent to $\frac{t^{s-1}}{{s-1!}} $. Remark
  that $S_{st}=\sum_{i=1}^{i=[p/2]}S_{s-1\ i}$. By induction on $s$,
  $S_{st}$ is equivalent to $\frac{t^{s-1}}{(s-1)!2^{\frac{s(s-1)}{2}}}$. 
\end{dem}

Summing up the situation, a number $q$ is a pseudo-key if the sum of the rests
$r_i(q)$ is less than $q$ and if these rests form a superincreasing
sequence. By proposition \ref{compteDesClesDePetitReste}, the
probability for the first condition is
less than $(\frac{3}{4})^{s-1}$. And by proposition \ref{nombreDeSuperCroissantes}, the
probability that the second condition is satisfied is around
$\frac{1}{2^{\frac{s(s-1)}{2}}} $. 

In particular we expect a unique pseudo key $q$ when 
the number of possible values for $q$ is asymptotically dominated by 
$(\frac{4}{3})^{s-1}2^{\frac{s(s-1)}{2}}$. This is the case for the 
second system we have constructed with the suggested choices of
parameters and this gives an explanation to the results of
proposition \ref{pseudoClesCompteExperimental}.

This is only a heuristic argument (there could be obvious pseudo-keys
associated to the private key $q$, for instance $q-1,q+1$ or $2q$). 
However, the general picture is that the 
unicity of the pseudo-key verified empirically in
proposition \ref{pseudoClesCompteExperimental}
should be  easy to reproduce with other families and other choices of
parameters.

\subsection{Finding a pseudo-key is as difficult as factorising an integer}
\label{sec:finding-pseudo-key}

In this section, we show 
that the problem of finding the exact value of the private key $q$ is as
difficult as factorizing a integer $n$, product of two primes. 
More precisely, we show that an easier problem (finding a pseudo-key
with the help of some extra-information ) is as difficult as the
factorisation of $n$, in the sense of a probabilistic reduction.

There are several problems, depending on whether one wants to compute
one key or all keys, and depending on the information given as input. 
\begin{listecompacte}
\item Input of problem 1: the public key $w_1,\dots,w_s$. Problem 1: compute all
  the pseudo-keys $q$ 
\item Input of problem 2: the public key $w_1,\dots,w_s$. Problem 2: compute one
  pseudo-key $q$ 
\item Input of problem 3: the public key $w_1,\dots,w_s$ and integers
  $r_1< \dots < r_{s-1}$, a range $[a,b]$. Problem 3: compute all 
  pseudo-keys $q$ such that the rests of the divisions $w_i=qv_i+\epsilon_i$,
  satisfy $\epsilon_i=r_i$ for $0<i<s$ and
  $\epsilon_s \in [a,b]$.   
\item Input of problem 4:  the public key $w_1,\dots,w_s$ and integers
  $r_1< \dots < r_{s-1}$, a range $[a,b]$. Problem 4: compute one
  pseudo-key $q$ such that the rests of the divisions $w_i=qv_i+\epsilon_i$,
  satisfy $\epsilon_i=r_i$ for $0<i<s$ and
  $\epsilon_s \in [a,b]$. 
\end{listecompacte}
Obviously, it is more difficult to find all the keys than to find one
key, and the problem is easier when more information is given as
input, as long as the definition of ``more difficult'' is sensible ( polynomial
time reduction, probabilistic polynomial time reduction ...).
In particular, if $>$ stands for ``more difficult'' then $problem\ 1>
problem\ 2$, and $problem\ 1>problem\ 3 > problem\ 4$ in the above
list. There is no proven relation beween $problem\ 2$ and $problem \
4$. However, when the
pseudo-key is unique, then $problem\ 1=problem\ 2$ 
and the easiest problem in the list is $Problem\
4$. The previous section explained why the pseudo-key is
unique for many cryptosystems. 
Thus the security of the system relies on the difficulty to solve 
$Problem\ 4$. We show that
solving $Problem\ 4$ is as difficult as factorising a product of two
primes. 
\begin{listecompacte}
\item Input of problem 5: an integer $n$ which is a product of two
  primes.  Problem 5: Find the factors $p,q$ of $n$.
\end{listecompacte}

\begin{thm}\label{trouverCle=factoriserUnNombre}
  If it is possible to solve $Problem\ 4$ in polynomial time
  (with respect to the length of the input data), then
  $\forall \eta >0$, it is possible to solve $Problem \ 5$ in
  polynomial time 
  with a probability of success at least
  $1-\eta$. 
\end{thm}
\begin{dem} Let $n$ be an integer.
  We make a polynomial time probabilistic reduction to $Problem\ 4$
to get the
  factorisation of $n=pq$.

Choose any superincreasing sequence
 $0<r_1<\dots < r_{s-1}$. 
First, try to divide $n$ by all elements
$q$ with $1<q \leq 3\sum_{i=1}^{s-1} {r_i}$. If this doesn't succeed, then all
 the divisors $q$ of $n$ satisfy $q>3\sum_{i=1}^{s-1}
 r_i$.

Let $w_i=n+r_i$ for $1\leq i
\leq s-1$. Let $r$ be an integer such that $(\frac{2}{3})^r<\eta$.
Let $w_{s1},\dots,w_{sr}$ 
be integers chosen randomly in the range
 $]\frac{n}{2} ,n[$. 
With these $r$ numbers, we consider $r$ problems $P_1,\dots,P_r$. 
The problem $P_k$ is $Problem \ 4$
with input
 $w_1,\dots,w_{s-1},w_{sk},r_1,\dots,r_{s-1},a=0, b=[\frac{n}{2} ]$. 

% Let $w_2=n+[\frac{\sqrt{n}}{4}]$, 
% $w_3=n+[\frac{\sqrt{n}}{2}]$.
% Let $w_{11},\dots,w_{1r}$ 
% be $r$ numbers chosen randomly in the range
%   $\{[\frac{n}{2} ],\dots,n-1\}$. 
% With these $r$ numbers, we consider $r$ problems $P_1,\dots,P_r$. 
% The problem $P_k$ is $Problem \ 4$
% with input
%   $w_{1k},w_2,w_3,\epsilon_2=[\frac{\sqrt{n}}{4}],
% \epsilon_{3}=[\frac{\sqrt{n}}{2}]$. 

% Let $q>p$ be the greatest divisor of $n=pq$.
% For each $k$, there is a
%   probability $x>\frac{1}{4} $ that $\epsilon_{1k}:=w_{1k} \mod\ q$ verifies
%   $\epsilon_{1k}<\frac{q}{4}$. 
%  Choose $r$ such that
%   $(1-x)^r<\eta$. Then, with probability at least
%   $(1-\eta)$, among the $r$ random choices 
% $w_{11},\dots,w_{1r}$  for $w_s$, one of them $w_{1k}$ 
% satisfies $\epsilon_{1k}<\frac{q}{4}$. 
% We denote by $(*)$ this condition. To conclude, 
% it suffices 
% to find a factorisation of $n$ in polynomial time
% when $(*)$ is satisfied. 

Let $q$ be a proper divisor of $n=pq$.
 It satisfies $q>3\sum_{i=1}^{i=s-1}
  r_i$. Thus, for each $k$, there is a
  probability $x>\frac{1}{3} $ that $w_{sk} \mod\ q$ satisfies
  $\sum_{i=1}^{s-1}r_i<w_{sk} \ mod \
  q<q$. Remark that
  $(1-x)^r<(\frac{2}{3} )^r<\eta$. Then, with probability at least
  $(1-\eta)$, among the $r$ random choices 
$w_{s1},\dots,w_{sr}$  for $w_s$, one of them $w_{sk}$ 
satisfies  $\sum_{i=1}^{s-1}r_i<w_{sk} \ mod \
  q<q$. 
We denote by $(*)$ this condition. To conclude, 
it suffices 
to show that one can find a factorisation of $n$ in polynomial time
when $(*)$ is satisfied.

We thus suppose that one problem $P_k$ in the list 
$P_1,\dots,P_r$ satisfies the condition $(*)$.
Since $r_i<q$, the equality $w_i=qp+r_i$ is the euclidean division of
$w_i$ by $q$ when $0<i<s$. Since the rest $\epsilon_{sk}$ of the division
$w_{sk}=q[w_{sk}/q]+\epsilon_{sk}$ satisfies
$\epsilon_{sk}>\sum_{i=1}^{s-1}r_i$ and $\epsilon_{sk}<q\leq
\frac{n}{2} $,  
it follows that a proper divisor $q$ 
of $n$ is a solution to problem $P_k$.

% We thus suppose that one problem $P_k$ in the list 
% $P_1,\dots,P_r$ satisfies the condition $(*)$.
% Since $\epsilon_2=[\frac{\sqrt{n}}{4}
% ]<\frac{q}{4} $, the equality $w_{2}=qp+\epsilon_{2}$ is the euclidean
% division of $w_2$ by $q$. Similarly, $w_{3}=qp+\epsilon_{3}$ is 
% an euclidean division. Since the rest $\epsilon_{1k}$ of the division
% $w_{1k}=q[w_{1k}/q]+\epsilon_{1k}$ satisfies
% $\epsilon_{1k}<\epsilon_2$, it follows that the greatest divisor $q$ 
% of $n$ is a solution to problem $P_k$. 

Reciprocally, a solution $q$ of $P_k$
  is a divisor of $n$ different from $1$ since $w_1 \mod q=r_1$. 
  This divisor of $n$ is not
  $n$ since the condition $\epsilon_{sk}\in [a,b]$ is not satisfied for
  $q=n$. 
  Thus a polynomial time algorithm that solves $Problem\
  4$ returns a strict divisor $q$ of $n$ when applied to
  $P_k$. Hence the
  factorisation of $n$ in polynomial time. 

A priori, we don't know which problem $P_k$ satisfies $(*)$ in the
list $P_1,\dots,P_r$. We thus run a multi-threaded algorithm which tries to
  solve in parallel the problems $P_1,\dots,P_r$
and which stops as soon as it finds a solution for one problem. 
\end{dem}

\subsection{Comparing LLL attacks on $x_0$ and $x_1$}
\label{sec:comp-lll-attacks}

The previous sections have explored the security of the key. 
It remains to analyse the security of the system with respect to 
heuristic attacks. As most heuristic attacks of knapsack cryptosystems
rely on variants
of the LLL algorithm, we analyse the security of the system for
LLL-based heuristic attacks. 

The knapsack problem 
is NP-complete and experiments show that the heuristic attacks fail
when the encryption is done with a well chosen general key $x_0$. 
In our system, the encryption is realised with a key
$x_1=qx_0+\epsilon$ which is a modification of $x_0$, and it could
happen that the key $x_1$  
is less secure than $x_0$. Thus we look for a security result
asserting that the key $x_1$ is as secure as $x_0$
for LLL-attacks.

The key $x_1$ could
be weaker than $x_0$ for two reasons:
\begin{listecompacte}
\item the heuristic algorithm used to break the system could perform faster for a message 
encrypted with 
$x_1$ than with a message encrypted with $x_0$
\item The heuristic could fail for a message encrypted with
$x_0$ but could succeed for the same message encrypted using $x_1$.
\end{listecompacte}

We fix an algorithm to attack the ciphertexts.
To measure the speed of the algorithm, we denote by $n(N)$ the 
number of steps of the
algorithm when the attack is run on the ciphertext $N$. 
To measure the probability of success of the algorithm, we introduce
the symbol $R(N)$ which is the result of the attack ( $R(N)=m$ if the
attack succeeds and recovers the plain text message $m$, $R(N)=FAILURE$ otherwise). 
As the algorithm depends on a matrix $M$ 
chosen randomly in the unit ball $B(1)$, the precise notations are
$n_M(N)$ and $R_M(N)$.

The two keys $x_0$ and $x_1$ yield two ciphertexts $N_0$
and $N_1$.  The following theorem says that the key
$x_1=qx_0+\epsilon$ is as secure as $x_0$ 
both from speed consideration and probability of success of the
attack. Both the numbers of steps $n$ and the returned message $R$ are
unchanged when replacing $x_0$ with $x_1$ 
provided that two conditions are satisfied: the matrix $M$
must live in a dense open subset and $\frac{||\epsilon||}{|q|} $ must 
be small enough. These two conditions 
are compatible with the practice:  $M$ is
chosen randomly and falls with high probability in a 
dense open subset and $\frac{||\epsilon||}{|q|} $ is small
by the very construction of our cryptosystem. 

\begin{thm}
  $\forall m, \forall x_0$, there exists a dense open subset $V\inc
  B(1)$, there exists $\eta>0$ such that  $\forall M\in V$, 
  $\forall x_1=qx_0+\epsilon$ with $\frac{||\epsilon||}{|q|} <\eta$:
\begin{listecompacte}
\item $n_M(N_0)=n_M(N_1)$
\item $R_M(N_0)=R_M(N_1)$. 
\end{listecompacte}
\end{thm}

The key arguments of our proof are as follows:
\begin{listecompacte}
\item The elements $x_1$ and $x_0$  are
  close as points of the projective space
\item The LLL algorithm can be factorized to give an action on the
  projective level
\item The number of steps in the algorithm and the result of the
  algorithm are functions of the input which are 
locally constant on a dense open subset.
In particular, replacing $x_0$ with $x_1$ does not change the number
of steps and the result when $x_0$ and $x_1$ are sufficiently close. 
\end{listecompacte}
 Though the algorithm required for the attack is fixed, its precise form is not
 important. The key point is that it relies 
on the LLL algorithm and that the additional data $M$ required to 
run the algorithm is chosen randomly. Similar
theorems can be obtained with other heuristics relying on the LLL
algorithm. Thus, besides the precise attack considered, our theorem
suggests that replacing the public key $x_0$ with $x_1$ does not
expose our system to LLL-based attacks.

\subsubsection{The LLL-algorithm}
\label{sec:lll-algorithm}
This section shows that the output of the LLL-algorithm 
depends continuously of the input when the input 
takes value in a dense open subset.

This is not clear a priori, since
the operations performed during the LLL algorithm
include non continuous functions ( integer
parts). We introduce a 
class of algorithms that we call 
 analytic.
The LLL algorithm is an analytic algorithm.
Analytic algorithms can include 
non continuous functions in the process but their
output depends continuously (in fact analytically) of the input
when the input is general enough.  

Recall that the LLL algorithm takes for input a basis $(b_1,\dots,b_n)$ of a
lattice $L\inc \RR^m$ and computes a reduced basis $(c_1,\dots,c_n)$. We refer 
to \cite{menezesVanooschotVanstone97:handbookCrypto}
 for details.

\begin{defi}\label{def:algoLocAnalytique}
  Consider an algorithm which makes operations on a datum $D\in U$ where
  $U\inc \RR^n$ is an open set (each step of the algorithm is a
  modification of the value of the datum $D$). Suppose that the algorithm
  is defined by a number of states $0,1,\dots,s$ and
  for each state $i$ by:
  \begin{listecompacte}\item a function $f_i:U\fd \RR$ 
  \item two functions $T_i^+:U\fd U$ and $T_i^-:U\fd U$ 
  \item two integers $i^+$ and $i^-$ in $\{0,\dots,s\}$. 
  \end{listecompacte}
The algorithm starts in state 1 with datum $D$ the input of the
algorithm. If the algorithm is in state $i$, 
the datum is $D$ and $f_i(D)>0$ (resp. $f_i(D)\leq 0$), then 
it goes to state $i^+$
(resp. $i^-$) with the datum $T_i^+(D)$ (resp. $T_i^-(D)$). 
The algorithm terminates in state $0$ and returns the value of the
datum $D$ when it terminates. By convention, we put
$0^+=0^-=0$, $T_0^+=T_0^-=Identity_U$, $f_0=1$.  

The algorithm is called  analytic if:
\begin{listecompacte}
\item  the test functions $f_i:U\fd \RR$ are analytic
  \item the transformation functions $T_i^+:U\fd U$ and $T_i^-:U\fd U$ are analytic on a
    dense open subset $U_i\inc U$ such that $V_i=U\setminus U_i$ is a
    closed analytic subset
  \item For every $D$ in $U$, the algorithm terminates.
\end{listecompacte}
\end{defi}

\begin{prop}\label{LLLestAnalytique}
  The LLL alogorithm is  analytic. 
\end{prop}
\begin{dem}
  We use the description of the algorithm described in
  \cite{menezesVanooschotVanstone97:handbookCrypto}, page
  119. The datum $D$ handled by the algorithm 
  is a basis $(b_1,\dots,b_n)$ of a lattice $L$. It takes
  values in the open subset $U\inc (\RR^m)^n$ parametrising
  the $n$-tuples of linearly independent vectors.
  All the tests functions $f_i$ which appear in
  the algorithm of \cite{menezesVanooschotVanstone97:handbookCrypto}
  are analytic (they are even
  algebraic functions on $U$). All the functions 
  involved in the
  handling of the basis $b_i$ (which correspond to our functions 
 $T_i^+$ and $T_i^-$) are algebraic too, except for an integer
  part $[x]$ which is analytic on the dense open set $x\notin \NN$. 
\end{dem}

%Since the LLL algorithm is  analytic, the following theorem
%obviously implies proposition \ref{LLLMarcheBien} formulated at
%the beginning of the section.

\begin{thm}\label{thmAlgoAnalytique}
  Let $A:U \fd U$ be the output function associated to 
  an  analytic algorithm ie. for $D\in U$, the value of $A(D)$ is the
  output of an analytic algorithm with input $D$. Then there exists a
  dense open subset $V\inc U$ such that
  \begin{listecompacte}\item 
    $A:V\fd U$ is analytic
  \item the number of steps to
    compute the output $A(D)$ is locally constant for $D\in V$.
  \end{listecompacte}

\end{thm}
\begin{dem}
  We keep the notations of definition \ref{def:algoLocAnalytique}. 
  In particular, the
  algorithm starts in state $1$ and ends in state $0$. 
  A sign function $\epsilon$ of length $length(\epsilon)=k$ is by definition a function 
  $\epsilon:\{1,\dots,k\}\mapsto \{+,-\}$. We associate to any sign
  function of length $k$ a finite sequence
  $n_0(\epsilon),\dots,n_{k}(\epsilon)$ constructed with the integers
  $i^+$ and $i^-$ of the analytic algorithm. Explicitly 
  $n_0(\epsilon)=1$,
  $n_1(\epsilon)=n_0(\epsilon)^{\epsilon(1)}$, \dots,
  $n_k(\epsilon)=n_{k-1}(\epsilon)^{\epsilon(k)}$. 
  We use below the notation $n_i$ instead of $n_i(\epsilon)$ to
  shorten the notation. Let
  $A_\epsilon:U\fd U$, $A_\epsilon=T_{n_{k-1}}^{\epsilon(k)}\circ \dots \circ
  T_{n_1}^{\epsilon(2)}\circ T_{n_0}^{\epsilon(1)}$.  
Let $g_\epsilon:U\fd \RR$, $g_{\epsilon}=f_{n_k}\circ A_\epsilon$.  
We define by
  induction on $k=length(\epsilon)$ a set $W_{\epsilon}$ such that 
  \begin{listecompacte}
  \item $W_{\epsilon}\inc U$ is an open inclusion
  \item $A_\epsilon:W_{\epsilon} \fd U$ is analytic.
  \item $D\in W_{\epsilon} \Rightarrow$  the successive states
    $s_0,\dots,s_k$ of the
    algorithm $A$ applied with input $D$ are $s_0=n_0(\epsilon)=1$, 
    $s_1=n_1(\epsilon)$,\dots,$s_k=n_k(\epsilon)$. Moreover, the value
    of the datum after the algorithm arrives in state $n_k(\epsilon)$ 
    is $A_\epsilon(D)$.  
  \item $\cup_{length(\epsilon)=k} W_\epsilon$ is dense in $U$. 
  \end{listecompacte}
We start the induction with $k=0$, using the convention that there is
a unique function $\epsilon$ defined on a set with $k=0$ element
and that $A_\epsilon=Id$. 
Then $W_\epsilon=U$ obviously satisfies the list of required
conditions.  

Let now $k>0$. Let 
$\tau:\{1,\dots,k-1\}\mapsto \{+,-\}$
be the restriction of $\epsilon$ to $ \{1,\dots,k-1\}$.
 
Let $W_{\tau+}=W_{\tau}\cap  \{D\in U, g_\tau(D)>0\} \cap
  (A_\tau)^{-1}(U_{n_{k-1}})$ where $U_{n_{k-1}}$ is the open subset of $U$ where
  $T_{n_{k-1}}^+$ and $T_{n_{k-1}}^-$ are analytic.
  Similarly, let  $W_{\tau-}=W_{\tau}\cap  \{D\in U, g_\tau(D)<0\} \cap
  (A_\tau)^{-1}(U_{n_{k-1}})$. The disjoint union $ W_{\tau+} \coprod W_{\tau-}$
    is dense in $W_{\tau}$ since the difference is included in the closed analytic
    subset $(g_\tau=0) \cup A_\tau^{-1}(U-U_{n_{k-1}})$.

Let $W_\epsilon=W_{\tau +}$ if $\epsilon(k)=+$ and
$W_{\epsilon}=W_{\tau -}$ if $\epsilon(k)=-$. Since 
 $ W_{\tau+} \cup W_{\tau-}$
    is dense in $W_{\tau}$ and since  $\cup_{length(\tau)=k-1}
    W_\tau$ is dense in $U$ by induction, we obtain the density of 
  $\cup_{length(\epsilon)=k} W_\epsilon$ in $U$. 

The other claims of the list are satisfied by construction. 

Let $W_k=\cup_{\epsilon\ of\ length\ k}W_\epsilon$. The intersection
$V=\cap_{k\geq 0}W_k$ is equal to the disjoint union 
$$\coprod _{k, \epsilon,length(\epsilon)=k, \ n_k=0,
  n_{k-1}\neq 0} W_{\epsilon}.$$ The set $V$ is open as a union of
open sets, and it is dense
in $U$ 
by Baire's theorem. On each open subset $W_\epsilon$ appearing in the disjoint
union, the algorithm applied to $D$ returns $A_\epsilon(D)$ which is
analytic and the number of steps of the algorithm is
$length(\epsilon)$, 
thus it is constant on each open set of the disjoint union.
\end{dem}

\begin{prop}\label{LLLMarcheBien}
  Let $b_1,\dots,b_n$ be a basis of a lattice $L\inc \RR^m$, $m\geq
  n$. Let $(c_1,\dots,c_n)=LLL(b_1,\dots,b_n)$ be the reduced basis
  computed by the 
  $LLL$ algorithm. There exists a
  dense open subset $U\inc (\RR^m)^n$ such that
  \begin{listecompacte}
  \item $U\mapsto (\RR^m)^n$, $(b_i)\mapsto (c_i)$ is continuous.
  \item $U\fd \NN$, $(b_i) \mapsto $number of steps of the
    $LLL$-algorithm is locally constant. 
  \end{listecompacte}
\end{prop}
\begin{dem}
  Follows from proposition \ref{LLLestAnalytique} and theorem \ref{thmAlgoAnalytique}.
\end{dem}

\begin{coro} \label{coro:changementBaseLLLContinu}
  Let $\psi:U\fd SL_{n}(\ZZ)$, $(b_1,\dots,b_n) \mapsto M$ such that 
$\left (
  \begin{array}{c}
    c_1 \\
    \dots \\
    c_n 
  \end{array}
\right)
=M 
\left (
  \begin{array}{c}
    b_1 \\
    \dots \\
    b_n 
  \end{array}
\right)
$ is locally
  constant. 
\end{coro}
\begin{dem}
  The map is continuous with values a discrete set. 
\end{dem}

\subsubsection{The heuristic attack}
\label{sec:heuristic-attack}

% Suppose that the attacker uses the following strategy.
% He knows the public key $x_1=(x_{11},\dots,x_{1s})$ and the encyphered text $v=x_1m$. 
% He proceeds heuristically. Let $y=(y_1,\dots,y_s)\in \ZZ^s$ such that
% $\sum x_{1i} y_i=v$. Let $\underline y=(y_1-\frac{1}{2},\dots,y_s-\frac{1}{2})$. 
% Let $L\inc \ZZ^s$ the lattice defined by $L=\{l\in \ZZ^s, x_1l=0\}$.  
% Then $d=y-m^t$ is the closest point to $\underline y$ in
% $L$. 

% Let $b_1,\dots,b_s$
% be a basis of $L$. Let $M\inc \ZZ^s\oplus \ZZ$ generated by $(b_i,0)$
% and $(\underline y,1)$. It is expected that $(\underline y-d,1)$ is the shortest vector of
% $M$. Since the first vector $z_1=(w_1,t_1)$ in the reduced $LLL$-basis is quite often
% the shortest vector of the lattice, a heuristic attack is as follows.
% \begin{listecompacte}
% \item Compute a base $B=(b_1,\dots,b_s)$ of $L$.
% \item Compute $y$
% \item Apply the $LLL$-algorithm to the family $(b_i,0),(y,1)$.
% \item Check wether the message $m$ is $w_1$. 
% \end{listecompacte}
% Though $B$ and $y$ are not uniquely determined, they depend on $x_1$
% and we denote them by $B(x_1)$ and $y(x_1)$. Instead of cyphering the
% message $m$ with $x_1$, we could have cyphered it with $x_0$. The
% following theorem says that if $x_0$ is very general, one can 
% construct from $x_0$ a public key $x_1$ such that uncyphering
% a message encyphered with $x_1$ is as difficult as uncyphering a
% message encyphered with $x_0$. 

Let $w_1,\dots,w_s \in \NN$ be a public key. Let $m\in \{0,1\}^s$ be a
plaintext message and
$N=\sum_{i=1}^s m_iw_i$ be the associated ciphertext. 
The following attack is well known. 
\\
\textbf{Heuristic Attack 1.}
\begin{listecompacte}
\item Choose $\lambda=2^{-2s}min(w_i)$
\item Apply the LLL algorithm to the lattice generated by the rows
  $b_i$ of
  the matrix $D=\left( 
  \begin{array}{ccccc}
    \lambda & 0     & \dots & 0     & w_1 \\
    0 & \lambda     & \dots & 0     & w_2 \\
\dots & \dots & \dots & \dots & \dots \\
    0 & 0     & \dots & \lambda     & w_s \\
    0 & 0     & 0     & 0     & N 
  \end{array} \right)$. 
Any vector $c_i$ of the reduced basis is a linear combination: 
 $c_i=\sum_{j=1}^{j=s+1} r_{ij} b_j$ 
\item For each vector $c_i$ of the reduced basis, check if
  the set ${r_{ij}}, j\leq s$ (or $-{r_{ij}}$) is equal to $m$ (ie. 
  check if $r_{ij}=0$ or $1$,  and if $\sum_{j=1}^{j=s}
  r_{ij}w_j=N$)
\end{listecompacte}

In the above attack, the precise value of the coefficients of the matrix $D$ 
is not important. The precise shape of $D$ has been chosen to speed-up
the computations and simplify the presentation, but is not required by
theoretical considerations. The attack could start with 
any invertible matrix  whose $s$ first columns 
contain small numbers and 
whose last column is close to the last column of $D$. 
Thus the following attack
is more general and natural.\\ 
\textbf{Heuristic attack 2.}
\begin{listecompacte}
\item Choose $\lambda=2^{-2s}min(w_i)$
\item Choose
  coefficients $m_{ij},i,j\leq s+1$ with $|m_{ij}|\leq 1 $. Let $M=(m_{ij})$ be
  the corresponding matrix.
\item Let $X=\left( 
  \begin{array}{ccccc}
    0 & 0     & \dots & 0     & w_1 \\
    0 & 0     & \dots & 0     & w_2 \\
\dots & \dots & \dots & \dots & \dots \\
    0 & 0     & \dots & 0     & w_s \\
    0 & 0     & 0     & 0     & N 
  \end{array} \right)$. 
Apply the LLL algorithm to the lattice generated by the rows
  $b_i$ of
  the matrix 
\begin{displaymath}
D=X+\lambda M=\left( 
  \begin{array}{ccccc}
    \lambda m_{11} &      & \dots & \lambda m_{1s}     & w_1+\lambda m_{1,s+1} \\
    \lambda m_{21} &     & \dots & \lambda m_{2s}     & w_2+\lambda m_{2,s+1} \\
\dots & \dots & \dots & \dots & \dots \\
    \lambda m_{s1} &      & \dots & \lambda m_{ss}     & w_s+\lambda m_{s,s+1} \\
    \lambda m_{s+1,1} &      &      & \lambda m_{s+1,s}     & N+\lambda m_{s+1,s+1} 
  \end{array} \right).
\end{displaymath}
Any vector $c_i$ of the reduced basis is a linear combination: 
 $c_i=\sum_{j=1}^{j=s+1} r_{ij} b_j$ and the coefficients $r_{ij}$ can be computed
 during the LLL algorithm.
\item For each vector $c_i$ of the reduced basis, check if
  the set ${r_{ij}}, j\leq s$ or $-{r_{ij}}, j\leq s$ is equal to $m$.
\end{listecompacte}

\subsubsection{Proof of the theorem}
\label{sec:proof-theorem}

Consider a plain text message $m$. It can be encrypted with the generic key
$x_0=(v_1,\dots,v_s)$
or with the key $x_1=qx_0+\epsilon=(w_1,\dots,w_s)$.  The two ciphertexts
associated with the keys $x_0$ and $x_1$ are denoted by $N_0$ and
$N_1$.  

We compare below how these two encryptions resist to ``Heuristic attack
2'' presented above. For this algorithm, we need a random matrix $M$
in the unit ball $B(1)$. Recall that we called $n_M(N)$ the number of 
steps of the algorithm when the attack is done on the ciphertext $N$. 
Similarly, we defined $R_M(N)$ to be the result of the attack
($R_M(N)=m$ if the attack recovers the plain text message $m$ and
$R_M(N)=FAILURE$ otherwise). 

% Thus, to compare the speed
%  and the probability of success of the attacks for the two keys $x_0$
%  and $x_1$,  one needs to compare 
% \begin{listecompacte}\item 
%   $n_M(N_0)$ and $n_M(N_1)$
% \item 
%   $R_M(N_0)$ and $R_M(N_1)$
% \end{listecompacte}

% If one chooses the parameters
% $\lambda_0=\lambda,M_0=M$ to run the attack on $x_0$, it is sensible to take the
% coefficients $\lambda_1=q\lambda_0, M_1=M_0$ for $x_1$ (as $||x_1||_\infty\simeq
% q||x_0||\infty$) to compare both attacks. Thus, to compare the speed
% and the probability of success of the attacks on the ciphertexts
% $N(x_0,m)$ and $N(x_1,m)$, one needs to compare 

\begin{thm}\label{thm:comparaisonCleGeneriqueEtSpeciale}
  $\forall m, \forall x_0$, there exists a dense open subset $V\inc
  B(1)$, there exists $\eta>0$ such that  $\forall M\in V$, 
  $\forall x_1=qx_0+\epsilon$ with $\frac{||\epsilon||}{|q|} <\eta$:
\begin{listecompacte}
\item $n_M(N_0)=n_M(N_1)$
\item $R_M(N_0)=R_M(N_1)$. 
\end{listecompacte}
\end{thm}
\begin{dem}
We keep the notations $X,\lambda,D=X+\lambda M$
introduced in the description of the attack. These data depend on 
the public key $x=(w_i)$. We denote by $X_0,\lambda_0,D_0$ and
$X_1,\lambda_1,D_1$ these data for the keys $x_0$ and $x_1$.

If  $C(\epsilon,q)$ is the
matrix defined by 
$X_1=q(X_0+C(\epsilon,q))$, then   $C(\epsilon,q))\fd 0$
when  $\frac{||\epsilon||}{|q|}\fd 0$.

If $M$ is a matrix with lines $b_1,\dots,b_s$, and if
$(c_1,\dots,c_s)=LLL(b_1,\dots,b_s)$ is the reduced basis 
computed by the LLL-algorithm, we adopt a matrix notation and 
we denote by $LLL(M)$ the matrix 
with lines $c_1,\dots,c_s$. We denote by $\psi(M)$ the matrix 
that gives the base change ie. $LLL(M)=\psi(M).M$. Finally, we denote
by $n(M)$ the number of steps to perform the LLL-algorithm on the
lines of $M$. 

According to proposition \ref{LLLMarcheBien} and corollary 
\ref{coro:changementBaseLLLContinu}, there exists a
dense open subset  $U$ where LLL is continuous and where $n$ and
$\psi$ are locally constant. 

Let $V=\frac{U-X_0}{\lambda_0} \cap B(1)$. Thus $V$ is a dense
open subset in $B(1)$ where the map $\psi_0: M\mapsto \psi(D_0(M))$ is
continuous. Moreover, the number of steps of the algorithm which computes
$\psi_0$ is locally constant on $V$.   

The analysis of the LLL algorithm given in
\cite{menezesVanooschotVanstone97:handbookCrypto} 
shows that it is a
``projective algorithm'' ie, in symbols: 
  if $\rho\in \RR$, we have $LLL(\rho M)=\rho LLL(M)$,
  $\psi(\rho M)=\psi(M)$ and $n(\rho M)=n(M)$.

% Let $WW\in M_{n+1\x n+1}(\RR)$ be the matrix with coefficients $M_{ij}$
% and $X=M-WW$. Let $U$ be the dense open subset where the $LLL$-algorithm is
% continuous and $V=U-X$. For $M\in V$, $M+X\in U$, thus the map $M\fd
% LLL(M+X)$ is continuous on $U$. 

By definition of the attack considered, 
the result $R_M(N_i)$ of the attack is a function of the
coefficients $r_{ij}$ which appear in the matrix 
$\psi(D_i(M))$. In particular, if $\psi(D_0(M))=\psi(D_1(M))$, then 
$R_M(N_0)=R_M(N_1)$.

$\psi(D_1(M))=\psi(q(X_0+C(\epsilon,q))+\lambda_1M)
=\psi(X_0+C(\epsilon,q)+\frac{\lambda_1M}{q})
=\psi(X_0+\lambda_0(\frac{\lambda_1M}{q\lambda_0}+\frac{C(\epsilon,q)}{\lambda_0}
))=\psi_0(\frac{\lambda_1M}{q\lambda_0}+\frac{C(\epsilon,q)}{\lambda_0 })$.  
When $\frac{||\epsilon||}{|q|}\fd
0$, the argument of $\psi_0$ tends to $M$. 
Since $M$ is in the open set of continuity of $\psi_0$,
and since $\psi_0$ is locally constant, 
$\psi_0(\frac{\lambda_1M}{q\lambda_0}+\frac{C(\epsilon,q)}{\lambda_0
})
= \psi_0(M)=\psi(D_0(M))$ if  $\frac{||\epsilon||}{|q|}$ is small
enough. 

Since $n$ is locally constant too, one can do a 
similar reasoning with $n$ instead of $\psi$ to show that 
$n_M(N_0)=n(D_0(M))=n(D_1(M))=n_M(N_1)$. 
% Choose any $\epsilon_i$. When $c\fd \infty$, $cx_0+\epsilon$ tends to 
% infinity $x_0$ in the projective space. In particular $cM+X(c)$ tends
% to $M+X_0$ in the projective space, and for $c>>0$, it is in the same 
% connected component as $M+X_0$. Since $n$ and $R$ are constant on the 
% connected component,  $n(M,N(x_0,m))=n(M',N(x_s,m))$
% and $R(M,N(x_0,m))=R(M',N(x_s,m))$.
\end{dem}

 \bibliographystyle{plain} 

\end{document}